\documentstyle[preprint,aps]{revtex}

\begin{document}
\preprint{IUCM93-010}

\title{Addition Spectra of Quantum Dots in Strong Magnetic Fields}

\author{S.-R. Eric Yang\cite{byline} and A.H. MacDonald}
\address{Indiana University, Department of Physics, Bloomington IN  47405}
\author{M.D. Johnson}
\address{Department of Physics, University of Central Florida,
Orlando FL 32816-2385}

\date{\today}

\maketitle

\begin{abstract}

We consider the magnetic field dependence of the
chemical potential for parabolically
confined quantum dots in a strong magnetic field.
Approximate expressions based on the notion that the
size of a dot is determined by a competition between
confinement and interaction energies are shown to be
consistent with exact diagonalization studies for
small quantum dots.  Fine structure is present in the
magnetic field dependence which cannot be explained without
a full many-body description and is
associated with ground-state level
crossings as a function of confinement strength or
Zeeman interaction strength.  Some of this fine structure
is associated with precursors of the bulk incompressible
states responsible for the fractional quantum Hall effect.

\end{abstract}


\narrowtext

Advances in nanofabrication technology have made it
possible to realize artificial systems in which
electrons are confined to a small area within a
two dimensional electron gas.   Interest in these
`quantum dot' systems\cite{qdot} has been enhanced as a result
of recently developed techniques\cite{mceuen,ashoori}
which probe them spectroscopically.  The
quantity which is measured\cite{meir} in these experiments is
the magnetic field dependence of the `addition
spectrum', {\it i.e.}, the energy to add one electron to a dot.
This is given by $\mu_N \equiv E^0_N-E^0_{N-1}$ where
$E^0_N$ is the ground state energy of an $N$-electron dot.
Addition spectrum measurements have
generally been interpreted in terms of `constant interaction'
models in which electron-electron interactions within a
quantum dot are accounted for by including a charging energy
which is characterized by a fixed self-capacitance;
or, when this fails, by using Hartree or Hartree-Fock
approximations.  However, especially at strong magnetic fields,
quantum dots can have strongly correlated\cite{wagner,maksym}
ground states, some of which are precursors of the bulk
incompressible states responsible for the fractional quantum
Hall effect.  In this regime a complete interpretation of
addition spectra measurements requires an exact treatment
of the Coulombic electron-electron interactions.

In this Letter we report on numerical exact diagonalization
calculations of the addition spectrum for
quantum dots in a strong magnetic field.  We find
that the addition spectrum has a surprisingly rich magnetic
field dependence, showing
a large number of sharp features superimposed on a smooth background.
The smooth background can be accounted for using a simple
Hartree approximation.  The sharp features are associated
with energy-level crossings at fixed $N$, often between
strongly correlated states.
The role of the spin degree of freedom is non-trivial
and is not in general consistent with
expectations of exchange-enhanced spin-splitting
based on the Hartree-Fock approximation.
The constant interaction model fails qualitatively for
strong magnetic fields.

We consider a system of electrons in two dimensions (2D)
which are confined by a parabolic external potential\cite{note1},
$V(r)=m\Omega^2r^2/2$.  We confine our attention
here to the strong magnetic field limit\cite{note2},
$\Omega /\omega_c \geq 1$. ($\omega_c \equiv eB_{\perp}/mc$
is the cyclotron frequency and  $B_{\perp}$ is the
component of the magnetic field perpendicular to the 2D electron gas.)
In this limit\cite{qdot} the symmetric gauge
single-particle eigenstates
are conveniently classified by a Landau level index $n$
and an angular momentum index $ m=-n, \dots, \infty$, and
we can confine our attention to $n=0$.  The
single-particle orbitals in the lowest Landau level have energies
$\varepsilon_m=\hbar\omega_c/2 + \gamma (m + 1)$,
where $\gamma =m\Omega^2\ell^2 =\hbar\Omega^2/ \omega_c$ and
$\ell^2 \equiv \hbar c / e B_{\perp}$.
($\langle m |r^2 | m \rangle = 2 \ell^2 (m+1)$.)
The quantized kinetic energy of the lowest Landau level is a
constant which hereafter we absorb
into the zero of energy.  We use as the unit of energy the interaction
energy $e^2 / \epsilon \ell$.  Then the many-electron energies are
determined by two dimensionless numbers characterizing the ratio
of the confinement and Zeeman energies to the interaction energy;
respectively, $\tilde \gamma \equiv \gamma / (e^2 / \epsilon \ell)$
and $\tilde g \equiv g \mu B / (e^2 / \epsilon \ell)$.
Note that we explicitly include the possibility of tilted fields
since we believe that tilted-field experiments will prove to
be very valuable.  As discussed below we have evaluated the
ground state energy over a wide range of values for these two
parameters.

The Hamiltonian in our system is invariant under spatial rotations about
an axis perpendicular to the 2D plane and passing through the
center of the quantum dot, and also under rotations in spin space
about the magnetic field direction ($\hat \alpha$).
It follows that both
the total angular momentum $M_z$ and
$S_{\alpha} \equiv \vec S \cdot \hat \alpha$  are good quantum numbers.  It is
straightforward to choose a representation for the
many-body Hamiltonian which is diagonal in these two operators
and block diagonal for the Hamiltonian.
Eigenenergies may be expressed as a sum of interaction and
single-particle contributions,
\begin{equation}
E_i(N,M_z,S_{\alpha}) = U_i(N,M_z,S_{\alpha})
+ \gamma (N+M_z) - g\mu_BBS_{\alpha}.
\label{eq10}
\end{equation}
Here $i$ labels a state within a $(M_z,S_{\alpha})$ subspace,
and $U_i(N,M_z,S_{\alpha}) \propto e^2/ \epsilon \ell$ is determined
by exactly diagonalizing the electron-electron interaction term
in the Hamiltonian within this subspace\cite{longpaper}.
In our study we have used a
Lanczos algorithm to determine only the minimum interaction energy
within each subspace, $ U_0(N,M_z,S_{\alpha})$.  For
$N = 2,3,4,5,6$ we have considered all possible values of
$S_{\alpha}$, while for $N = 7,8$ we have considered only
fully spin polarized states with $S_{\alpha}=N/2$. In each case
we have considered all values of $M_z$ from the minimum value
consistent with the Pauli exclusion principle (see below) to
$M_z = 3 N (N-1) /2$, which is large enough to accommodate
an $m=3$ Laughlin droplet\cite{smitra,mjprl}.
For given values of $\tilde \gamma $ and $\tilde g$ the subspace
containing the ground state is determined by minimizing
$U_0(N,M_z,S_{\alpha}) + \tilde \gamma (N+M_z) - \tilde g S_{\alpha} $
over all values of
$M_z$ and $S_{\alpha}$ for which calculations have been performed.
This procedure results in a surprisingly rich phase diagram for
a quantum dot.

Results for $N=5$ and $N=6$ are shown in
Fig.~(\ref{fig1}) and Fig.~(\ref{fig2}).
Regions in the phase diagram are labeled by
by $(M_z,2S_{\alpha})$, the quantum numbers of
the state with lowest energy.  Along the boundary lines
in these phase diagrams ground state level crossings occur;
the slope of a line is given by
$(S_{\alpha} - S^{\prime}_{\alpha}) / (M_z - M'_z)$ and the intercept by
$(U_0(N,M'_z,S'_{\alpha})-U_0(N,M_z,S_{\alpha}))/(M_z-M'_z)$.
It follows from the spin-rotational invariance of the electron-electron
interaction term in the Hamiltonian that states may
be labeled by a total spin quantum number $S$ and by
$S_{\alpha} = -S, \cdots, S$.  In each spin multiplet the
only ground state candidate for any non-zero $\tilde g$
is the state which is polarized along the field,
{\it i.e.} $S_{\alpha} = S$.  Thus the $S_{\alpha}$ values
in these figures give the total spin quantum numbers of the
corresponding states.

We discuss these rather complicated phase diagrams, beginning
with $\tilde{g}$ and $\tilde{\gamma}$ relatively large, on the
upper right-hand side of the figures.
For $N=5$ only the (4,1),
(6,3), and (10,5) regions in the phase diagram correspond to
the single Slater determinant ground states which would
be obtained in the Hartree-Fock approximation.  The occupation numbers
for these states are given by
$(\bullet\bullet\bullet\circ ;\bullet\bullet\circ ),\;
(\bullet\bullet\bullet\bullet\circ ;\bullet\circ ),\;\hbox{ and }\;
(\bullet\bullet\bullet\bullet\bullet\circ ;\circ )$, respectively.
[An occupied(unoccupied) single-particle state is represented by a
full(empty) circle.  Circles left(right) of
the semicolon represent spin up(down) states.
The angular momentum $m$ of a single-particle state
increases from left to right.]
Similarly, for $N=6$ the (6,0),
(7,2), (10,4), and (15,6) regions have single Slater determinant
ground states with occupation numbers given by
$(\bullet\bullet\bullet\circ ;\bullet\bullet\bullet\circ ),\;
(\bullet\bullet\bullet\bullet\circ ;\bullet\bullet\circ ),\;
(\bullet\bullet\bullet\bullet\bullet\circ ;\bullet\circ ),\;\hbox{ and }\;
(\bullet\bullet\bullet\bullet\bullet\bullet\circ ;\circ )$
respectively.  The (4,1) state for $N=5$ and the
(6,0) state for $N=6$ minimize the confinement energy and
are ground states at all values of
$\tilde \gamma$ in the absence of electron-electron and
Zeeman interactions.
These states are the precursors of the Landau level filling
factor $\nu =2$ states for bulk systems.

As the confinement strength $\tilde{\gamma}$ weakens, interactions favor
less compact (larger total angular momentum) electron dots
[$U_0(N,M_z+1,S_{\alpha}) \le U_0(N,M,S_{\alpha})$].
For these dot sizes the expansion is first
accomplished, except at small $\tilde g$, by forming the
most compact states consistent with increasing spin polarization
until complete spin polarization is reached.
For large $\tilde g$, states with
large spin quantum numbers are favored; eventually, for very large $\tilde{g}$,
only states with $S=N/2$ occur.
The tendency toward complete
spin polarization is what simplifies the phase diagram at
larger values of $\tilde g$.  At small $\tilde g$, as
the confinement $\tilde{\gamma}$ weakens the dot expands by introducing
holes\cite{mjprl,ourajp} into the interior of the dot.
As these holes begin to correlate
the Hartree-Fock approximation begins to fail.  One consequence
is that interactions often favor states which are not completely
spin-polarized.  At weaker
confinement the ground states are linear combinations of many Slater
determinants.  Many of the states which occur can be identified
as precursors of the bulk incompressible states responsible for
the fractional quantum Hall effect.
For example for $N=5$ the (30,5) region corresponds to the
$\nu = \frac{1}{3}$
state while for $N=6$ the (36,0) and (45,6) phase regions correspond
to the $\nu = \frac{2}{5}$ spin-singlet state
and $\nu = \frac{1}{3}$ spin-polarized states.

States with larger
values of $M_z$ occur and the phase diagram becomes richer as
$\tilde \gamma$ decreases.  The upper panels in Fig.~(\ref{fig1})
and Fig.~(\ref{fig2}) show the small $\tilde{g}$, small $\tilde{\gamma}$
regions of the phase diagrams on an expanded
scale.  The dashed line shows the path taken through
these phase diagrams for a GaAs sample with
$\hbar \Omega = 2 {\rm meV}$ as a function of a perpendicular
magnetic field.  (For GaAs $\tilde \gamma \sim 0.131
(\hbar \Omega [{\rm meV}])^2 / (B_{\perp} [{\rm Tesla}])^{3/2} $
and $\tilde g \sim 0.0059 B [\rm{Tesla}]/ (B_{\perp}[{\rm Tesla}])^{1/2}$.)
Regions of the phase diagram to the right of this line could
be explored experimentally by using tilted magnetic fields.

Some qualitative features of these results can be understood
using a simple argument which considers the competition between
the Hartree energy and the confinement energy
of a quantum dot.  We assume that in the ground state
electrons occupy the $N_{\phi}$ smallest-$m$ orbitals
with approximately equal probability, leading to a charge
distribution which is approximately that of a uniform disk of
radius\cite{note3}
$ R = \ell \sqrt{2 N_{\phi}}$.
(For such a state $M_z \sim  N N_{\phi}/2$.  The maximum
value of $N/N_{\phi}$ allowed by the Pauli exclusion principle
is $1$ for spin-polarized states and $2$ for unpolarized states.)
For all but the
smallest dots the two largest contributions to the total
energy will be the Hartree energy,
\begin{equation}
E_H \sim {8  e^2 N^2 \over 3 \pi \epsilon R}
= {e^2 \over \epsilon \ell} {4 \sqrt{2} \over 3 \pi}
{ N^2 \over N_{\phi}^{1/2}},
\label{eqone}
\end{equation}
and the confinement energy,
\begin{equation}
E_C =  \gamma (M_z+N) \sim \gamma N N_{\phi}/2 .
\label{eqtwo}
\end{equation}
Corrections due to exchange and correlations (which reduce
the interaction energy below $E_H$) give a contribution
proportional to $N^{1}$ for large $N$ are relatively less important
for large dots.
The confinement energy favors compact dots with small values
of $N_{\phi}$ while the interaction energy favors expanded dots.
For a given value of $\tilde \gamma$ and $N$ the optimum dot
size can be determined by minimizing $E_H+E_C$ with respect to
$N_{\phi}$.  This gives
\begin{equation}
{N_{\phi} \over N} =\left({4\sqrt{2} \over 3\pi \tilde \gamma N^{1/2}}
\right)^{2/3},
\label{eqthree}
\end{equation}
\begin{equation}
E_H+E_C = {3 \over 2}
[ (e^2 / \epsilon \ell)^2 \gamma (4 \sqrt{2}/ 3 \pi)^2]^{1/3} N^{5/3},
\label{eqfour}
\end{equation}
and
\begin{equation}
\mu_N \sim { 5 \over 2}
[ (e^2 / \epsilon \ell)^2 \gamma (4 \sqrt{2}/ 3 \pi)^2]^{1/3}
N^{2/3}.
\label{eqfive}
\end{equation}
Note that in this approximation the energy and $\mu_N$ are
independent of magnetic field.
This result differs qualitatively from the constant
interaction model where $\mu_N$ would be the sum of an
interaction term proportional to $N$ and a single-particle term.
The difference here is due to the fact that the
size of the dot is not fixed but is determined
by a competition of interaction and single-particle terms.
Comparing with
Fig.~(\ref{fig1}) and Fig.~(\ref{fig2}) we see that the
values of the ground state angular momenta are reasonably
estimated\cite{mjprl}
by Eq.~(\ref{eqthree}) (using $M_z \sim NN_{\phi}/2$) even for $N=5$
and $N=6$.  (Overestimates are expected since correlations will
reduce the interaction energy cost of making the dots smaller.)
The above Hartree argument predicts $N_{\phi}$ or $M_z$
in the ground state.  In a Hartree-Fock generalization of this
argument the exchange energy would stabilize the state with the
largest spin polarization allowed for a given $M_z$ by the
Pauli exclusion principle.
Indeed the most compact fully spin-polarized state
($M_z = N (N-1) /2 ; S_{\alpha}=N/2$),
which is the precursor of the bulk $\nu =1$ state,
has a large range of stability in the phase diagrams of
Fig.~(\ref{fig1}) and Fig.~(\ref{fig2}).  However, as seen
most clearly in the upper panels, many
states with smaller values of $S_{\alpha}$ occur at larger $M_z$ where
full spin polarization is allowed.  This is in direct contradiction
with Hartree-Fock theory and is a result of correlations.

Fig.~(\ref{fig3}) shows\cite{available}
the magnetic field dependence of
$\mu_6$ for a GaAs sample with $\hbar \Omega = 2 {\rm meV}$.
(The inset shows results for $N=2,3,4,5,6$ on a wider
energy scale.)
The approximately $N^{2/3}$ dependence of $\mu_N$ at fixed field
and the weak magnetic-field dependence are
explained by Eq.~(\ref{eqfive}).
Similarly, in approximate agreement with Eq.~(\ref{eqthree}),
the angular momentum difference between the $N=5$ ground state
and the $N=6$ ground state increases from $5$ to $15$ in going
from the left- to right-hand sides of the curve.
However, the finer features
apparent in the plot of $\mu_6$ can be understood only by accounting
for the possibility of strong correlations
in the quantum dot and cannot be explained with Hartree-Fock
or similar approximations.  The apparently smooth curve for
$N=6$ in the inset can be seen to have a large number of
cusps due to ground state level crossings for either $N=5$
or $N=6$ quantum dots.
At a ground state level crossing $dE_0/dB$ must decrease.  It follows
that ground state level crossings in the $N-1$ and $N$
particle systems lead respectively to positive and negative
jump discontinuities in $d \mu_N / d B$ as seen in Fig.~(\ref{fig3}).
Note that unlike the prediction of an independent-particle
approximation\cite{mceuen,ashoori}, upward and downward pointing
cusps do not in general alternate.
At the left-hand side ($B \sim 2.5 {\rm Tesla}$) of this figure both the
$N=5$ and $N=6$ dots are in the $(M_z = N (N-1)/2; S_{\alpha} = N/2)$
maximum-density spin-polarized
single Slater determinant states, while at the right-hand side
($ B \sim 6 {\rm Tesla}$) both $N=5$ and $N=6$ dots are in
$(M_z = 3 N (N-1) /2; S_{\alpha} =N /2)$ states.  These
states are the precursors of the bulk $\nu=1$ and
$\nu=1/3$ incompressible states and the incompressibility
is reflected\cite{longpaper}
in the relative large regions of stability in the
phase diagrams.  The resulting `plateaus' in the addition
spectrum should be among the most visible features
experimentally.  Precursors of a Landau level filling factor
$\nu$ state will occur for $N/N_{\phi}=\nu$; it follows from
Eq.~(\ref{eqthree}) that for GaAs we can expect associated
features in the addition spectrum to occur for
$B[{\rm Tesla}] \sim  0.363 ( \hbar \Omega {\rm [meV]})^{4/3} N^{1/3} / \nu$.
Features identified with $\nu =2 $ in the
recent experiments of Ashoori et al.\cite{ashoori} seem to
follow this $N^{1/3}$ law rather well.  We believe that
the unidentified experimental features which appear at
approximately twice this field are associated with precursors
of the $\nu =1$ incompressible state which is stabilized primarily
be electron-electron interactions.  We predict that features
associated with precursors of fractional incompressible states
will appear at stronger fields and also, less visibly, at
intermediate fields.

This work was supported by the National Science
Foundation under grant DMR-9113911 and by the
UCF Division of Sponsored Research.  AHM acknowledges helpful
conversations with Ray Ashoori, Mark Kastner, Horst
Stormer and Karen Tevosyan.

\begin{figure}
\caption{Phase diagram for a $N=5$ parabolically confined
quantum dot.  Regions in the phase diagram are labeled by
the $M_z$ and $N_{\uparrow}-N_{\downarrow}$ values of the
ground state.  ($S_{\alpha} = (N_{\uparrow} - N_{\downarrow})/2$.)
The upper panel shows the rich behavior at weak confinement
which is related to the physics of the fractional quantum Hall
effect.  The dashed line shows the path in the phase diagram
followed by GaAs sample with $\hbar \Omega = 2 {\rm meV}$ and
a perpendicular magnetic field between $B = 2.5 {\rm Tesla}$ and
$B = 7 {\rm Tesla}$.}
\label{fig1}
\end{figure}

\begin{figure}
\caption{Phase diagram for a $N=6$ parabolically confined quantum dot.
Regions in the phase diagram are labeled as in Fig.~(1).}
\label{fig2}
\end{figure}

\begin{figure}
\caption{Magnetic field dependence of the $N=6$ addition spectrum
for a parabolically confined quantum dot with $\hbar \Omega = 2 {\rm meV}$.
The curve has a cusp whenever there is a
ground state level crossing for either the $N=5$ or
$N=6$ dot.  Curve segments between two upward tick marks are labeled
with the ground state quantum numbers $(M_z,2S_z)$ of the $N=6$ dot.
Segments between downward tick marks are labeled with the quantum numbers of
the $N=5$ dot.  The paths followed through the phase diagrams for
this model are indicated by the dashed lines in Fig.~(1) and
Fig.~(2).  The inset shows results
for $N=2,3,4,5,6$ on a wider energy scale.
The dashed lines in the inset result from the Coulomb blockade model,
with a phenomenological self-capacitance obtained by a fit to the
exact numerical results.
}
\label{fig3}
\end{figure}

\begin{references}
\bibitem{note1} For the small $N$ quantum dots considered
here the use of a parabolic confinement potential model
is well justified.  For quantum dots defined by gates
the potential from external charges will change in a
geometry-dependent way as the quantum dot is charged.  These
changes can often be accounted for in terms of phenomenological
capacitances.
\bibitem{longpaper} Technical details will be described in
a planned longer publication.
\bibitem{available} A code which generates addition spectra
from the many-electron eigenenergy data sets is available from
the authors for use in interpreting experimental results.
\end{references}
\end{document}